\documentclass[titlepage,11pt]{article}
\usepackage{amsfonts}
\usepackage{amssymb}
\usepackage{amsmath}
\usepackage{amsthm}
\usepackage{thmtools}
\usepackage{mathtools}
\usepackage[doublespacing]{setspace}
\usepackage{natbib}
\usepackage{amssymb}
\usepackage{latexsym}
\usepackage{appendix}
\usepackage{float}
\usepackage{graphicx}
\usepackage{footmisc}

\begin{document}

\title{Propaganda, Alternative Media, and \\
Accountability in Fragile Democracies}
\author{Anqi Li\thanks{%
Department of Economics, Washington University in St. Louis.
anqili@wustl.edu.}, Davin Raiha\thanks{%
Kelley School of Business, Indiana University. draiha@iu.edu.}, and Kenneth
W. Shotts\thanks{%
Stanford Graduate School of Business. kshotts@stanford.edu.}}
\date{Forthcoming, Journal of Politics}
\maketitle

\begin{abstract}
We develop a model of electoral accountability with mainstream and
alternative media. In addition to regular high- and low-competence types,
the incumbent may be an aspiring autocrat who controls the mainstream media
and will subvert democracy if retained in office. A truthful alternative
media can help voters identify and remove these subversive types while
re-electing competent leaders. A malicious alternative media, in contrast,
spreads false accusations about the incumbent and demotivates policy effort.
If the alternative media is very likely be malicious and hence is
unreliable, voters ignore it and use only the mainstream media to hold
regular incumbents accountable, leaving aspiring autocrats to win
re-election via propaganda that portrays them as effective policymakers.
When the alternative media's reliability is intermediate, voters heed its
warnings about subversive incumbents, but the prospect of being falsely
accused demotivates effort by regular incumbents and electoral
accountability breaks down.

\bigskip


\bigskip

\bigskip

\bigskip

\bigskip

\bigskip

\bigskip

\bigskip

\bigskip

\bigskip

\bigskip

\bigskip

\bigskip
\bigskip
\bigskip
\bigskip
\bigskip

\bigskip
\bigskip
\bigskip
\bigskip
\bigskip
\bigskip

\bigskip
\bigskip
\bigskip
\bigskip
\bigskip
\bigskip

\bigskip
\bigskip
\bigskip

\noindent Supplementary material for this article is available in the online appendix in the online edition.
\end{abstract}

\noindent Many countries inhabit a grey area between democracy and
autocracy: their leaders are elected, but try to eliminate checks on their
power and subvert the institutional foundations of democracy. One check that
aspiring autocrats often remove is the mainstream media, which can be
induced, by a combination of censorship, ownership, and corruption, to
refrain from criticism and act as a propaganda vehicle for the regime.
Recent rulers who have taken this approach include Turkey's Erdo\u{g}an,
Hungary's Orban, and Venezuela's Ch\'{a}vez.

Citizens who are unsure about their leaders and skeptical of the mainstream
media may turn to alternative media.\footnote{%
Though the terms \textquotedblleft alternative media\textquotedblright\ and
\textquotedblleft independent media\textquotedblright\ are often used
interchangeably, we will use \textquotedblleft alternative
media\textquotedblright\ throughout the paper.} For our analysis, the
alternative media has two defining characteristics. First, it is independent
of the government and beyond its control. Second, citizens are unsure about
its intentions, which can range from providing truthful warnings about
aspiring autocrats to making malicious accusations against legitimate
leaders. The first kind of intention was demonstrated by Peru's Canal N,
which played a key role in the downfall of President Fujimori. The second
kind of intention was demonstrated by right-wing commentators in the United
States, who accused President Obama of many things, including being a
Marxist and, in the words of InfoWars's Alex Jones, a \textquotedblleft
would-be dictator.\textquotedblright\ Other examples of alternative media
that are difficult for the government to control include opposition
television and newspapers; foreign news providers; and, more recently,
social media platforms that contain a mixture of accurate information,
conspiracy theories, and disinformation. A key characteristic of all these
examples is that they can be either truthful or malicious and their
intentions are uncertain from (at least) some voters' perspectives.

We develop a model of electoral selection and accountability, in which
citizens are uncertain about the incumbent's type, the mainstream media's
independence, and the alternative media's intentions. Voters want to
incentivize and retain competent leaders while removing those who are
incompetent or autocratic. When the mainstream media praises the incumbent,
voters don't know whether this is neutral praise of skillful policymaking or
propaganda for an aspiring autocrat. The alternative media can be either a
truthful type that provides accurate warnings about autocratic subversion
and propaganda, or it can be a malicious type that makes false accusations.

We study how the alternative media affects accountability and selection. A
key parameter in our analysis is the alternative media's reliability,
defined as the probability that it is truthful. A highly-reliable
alternative media is beneficial, because it helps citizens remove aspiring
autocrats while holding regular non-autocratic politicians accountable for
their policymaking. On the other hand, if the alternative media is highly
unreliable, voters expect it to \textquotedblleft cry
wolf,\textquotedblright\ so they ignore its warnings. They use the
mainstream media to hold regular incumbents accountable, leaving autocratic
types to win re-election via propaganda and subsequently subvert democracy.

A more subtle effect occurs when the alternative media is too reliable to be
ignored by voters, yet sufficiently unreliable that competent incumbents
worry about being falsely accused. In that case, accountability breaks down
and selection is based only on the alternative media's report.

Our model has implications for two important features of contemporary
politics:\ fake news and democratic backsliding. Our model shows that fake
news doesn't just induce voters to make mistakes. Rather, false accusations
made by somewhat-reliable alternative media sources can also demotivate
incumbent effort and thereby undermine electoral accountability.

In the context of fragile democracies, our theory suggests a tension between
accountability for policymaking and prevention of democratic backsliding. In
many circumstances, elections cannot simultaneously achieve both of these
important goals, and voters who seek to re-elect effective leaders may fail
to heed warning signs that the incumbent is an aspiring autocrat.

Our theory sits at the intersection of the literatures on media bias,
electoral accountability, and fragile democracies. The media's role in
accountability has been examined by many scholars, including some who
analyze pro- or anti-incumbent biases (Ashworth and Shotts 2010, Warren
2012, Wolton 2019). While existing models (e.g., Besley and Prat 2006)
analyze capture of mainstream media, we study the effect of alternative
media that is independent of government control and either truthfully
reveals or falsely accuses the incumbent of being a subversive autocrat---a
feature that is absent from existing models.

The literature on autocracy and fragile democracy includes many theories of
propaganda and censorship (Egorov, Guriev, and Sonin 2009; Gehlbach and
Sonin 2014; Lorentzen 2014; Cheah 2016; Little 2017; Horz 2018). There is a
growing formal literature on democratic backsliding (e.g., Luo and
Przeworski 2019, Svolik 2020), but very few scholars analyze accountability
in weak democracies (though see Svolik 2013). Also, with the exception of
Nalepa, Vanberg, and Chiopris (2018) most models of backsliding don't
incorporate uncertainty about whether the incumbent is a subversive
autocrat. Our setting is closest to Guriev and Treisman's (2018) model of
propaganda and censorship by a regime that wishes to appear competent.
However, in their model citizens are only concerned about whether the ruler
is competent, and criticism of the incumbent is always accurate, whereas a
key parameter in our analysis is the reliability of the alternative media.%

\section*{Model Setup}

Consider a model of policymaking and elections, with five
actors: the incumbent, challenger, mainstream media, alternative media, and
voter. There are two equally-likely states of the world, $\omega \in \left\{
0,1\right\} $. The incumbent may learn $\omega $ and chooses a policy $x\in
\left\{ 0,1\right\} $, where the correct policy for the voter's interest is
the one that matches the state. Then the mainstream media announces a
message $m\in \left\{ 0,1\right\} $ about the policy it believes is correct.
The alternative media simultaneously issues a report $r\in \left\{
S,NS\right\} $ about whether the incumbent is a subversive type who uses the
mainstream media for propaganda. Finally, the voter observes the policy
choice $x$ and the mainstream and alternative media messages $m$ and $r$,
but not the true state $\omega $, and decides between the incumbent and
challenger.

The incumbent can be a high, low, or subversive type, $\theta _{I}\in
\left\{ H,L,S\right\} .$ High- and low-competence types are standard in the
accountability literature. The subversive type is novel: she controls the
mainstream media and if re-elected will consolidate her power and deliver a
negative payoff to the voter. The incumbent is subversive with probability $%
\sigma $. Conditional on not being subversive, she is a high type with
probability $\pi $. Low and subversive types only know the prior $\Pr \left(
\omega =1\right) =\frac{1}{2}$ and cannot acquire additional information. A
high type can exert effort, at cost $k$, to learn the true state before
choosing policy. The winner of the election gets an ego-rent of $1$.

The mainstream media is either truthful or propagandist. If truthful, it
non-strategically issues a report $m\in \left\{ 0,1\right\} $ that matches
the true state $\omega $ with probability $q\in \left( \frac{1}{2},1\right)
. $ The mainstream media is a propagandist if and only if the incumbent is
subversive, in which case it always reports that the incumbent's policy
choice was correct, $m=x$.

The alternative media is either truthful or malicious. If truthful, it
reports that the incumbent is subversive and is using the mainstream media
for propaganda, $r=S$, if and only if this is indeed the case. If malicious,
it always reports $r=S$.\footnote{%
Malice could be driven by either supply side factors (e.g., an owner who
wants to discredit the incumbent and the mainstream media) or demand side
factors (e.g., catering to a subset of the population that despises the
incumbent). Our results hold regardless of the source of malice.} A central
parameter in our analysis is the probability that it is malicious, $\phi \in %
\left[ 0,1\right] $. Hereafter $1-\phi $ is referred to as the alternative
media's \emph{reliability}.

After observing $x$, $m$ and $r$, the voter elects the incumbent or the
challenger. She gets utility $1$ from re-electing a high-type incumbent, $0$
from a low type, and $-s$ from a subversive type.\footnote{%
A possible objection is that some autocrats turn out to be effective
policymakers who are well-liked by their subjects. In our model, $-s$
represents voters' \emph{expected} payoff from retaining a subversive type.
This can include autocrats who turn out to be \textquotedblleft
good\textquotedblright\ along with those who turn out to be disastrous, as
long as the expected payoff is negative. We also note that even if the
autocratic type could exert high effort at a cost, she would have no
incentive to do so, because she can induce the mainstream media to praise
her policy decisions.} As is standard in accountability models, we assume
that the challenger is inactive until he is elected, in which case he
delivers an exogenous expected utility $U_{C}\in \lbrack -s,1]$ to the voter.

We characterize Perfect Bayesian equilibria that are symmetric with respect
to policies $0$ and $1$, which means low- and subversive-type incumbents
choose each policy with probability $\frac{1}{2}$. Other equilibrium
components are: (1) the high-type incumbent's effort decision and (2) the
voter's belief about the incumbent, as well as his election decision. In the
paper we focus on intuition; statements of equilibria can be found in the online appendix.\footnote{%
In the online appendix, we also show that our main results hold
qualitatively in an extended model, in which the subversive type decides
whether to capture the mainstream media and also has some ability to
influence the alternative media. Possible topics for future extensions
include allowing voters to sometimes directly learn $\omega $, incorporating
politicians with biased policy preferences, and making the model dynamic,
but a comprehensive analysis of such model variants is beyond the scope of
the current paper.}%

\section*{Baseline}

To establish a baseline, suppose there is no alternative
media. We say there is \emph{accountability} if two conditions hold: (1)
high-type incumbents exert effort to choose policy in the voter's interest
and (2) the incumbent is removed from office if $m\neq x$. These conditions
are mutually reinforcing: a high type exerts effort to choose good policies
and earn praise from the mainstream media, and the fact that she exerts
effort means the media message $m$ is informative about the incumbent's
type. We analyze whether it is possible to have accountability in
equilibrium, focusing on the incumbent's incentives to exert effort and the
voter's use of information provided by the media.

The voter's inferences about the incumbent are complicated by the
possibility of propaganda. If propaganda is very likely, the mainstream
media message is essentially meaningless. If propaganda isn't very likely, a
message that the incumbent chose the correct policy $(m=x)$ conveys positive
information about the incumbent's desirability. Assuming accountability,
high-type incumbents choose $x=\omega $ and low and subversive types choose
each policy with probability $\frac{1}{2}$, so the voter's expected utility
from re-electing the incumbent is:
\begin{equation*}
\overline{U}\coloneqq1\cdot \Pr \left( \theta _{I}=H|m=x\right) -s\cdot \Pr
\left( \theta _{I}=S|m=x\right) =\frac{\pi q-sl}{\pi q+\left( 1-\pi \right) 
\frac{1}{2}+l},
\end{equation*}%
where $l\coloneqq\frac{\sigma }{1-\sigma }$ is the likelihood that the
incumbent is subversive. Meanwhile, when $m\neq x$, the voter knows the
incumbent isn't subversive, and his utility from re-electing her is:
\begin{equation*}
\underline{U}\coloneqq1\cdot \Pr \left( \theta _{I}=H|m\neq x\right) =\frac{%
\pi \left( 1-q\right) }{\pi \left( 1-q\right) +\left( 1-\pi \right) \frac{1}{%
2}}.
\end{equation*}%
Accountability thus requires two conditions. First, the incumbent and
challenger must be similarly appealing ex ante, $U_{C}\in \lbrack \underline{%
U},\overline{U})$, so the voter re-elects the incumbent if and only if the
mainstream media reports $m=x$. A presumption of this is $\underline{U}<%
\overline{U}$, which, for any given $s>0$, holds when $l$ is sufficiently
low, i.e., subversive types aren't too likely. Second, the effort cost must
be sufficiently low$,$ $k\leq q-\frac{1}{2}$, so a high-type incumbent is
willing to exert effort to increase her probability of winning re-election
from $\frac{1}{2}$ to $q.$ Proposition 1 in the online appendix summarizes
equilibria for the baseline model.

\section*{Alternative Media}

We now analyze the full model, in which the alternative
media reports on whether the incumbent is subversive. To assess the effects
of the alternative media, we ask two questions. First, how does it affect
the incumbent's policymaking effort? Second, how does it affect electoral
selection, both in the sense of re-electing high-type incumbents and in the
sense of removing subversive ones?

\paragraph{Accountability}

The alternative media's reliability plays a key role in determining whether
accountability is possible. To see this, we begin with situations in which
accountability is possible in the baseline model, $U_{C}\in \lbrack 
\underline{U},\overline{U})$ and $k\leq q-\frac{1}{2}.$

We first consider extreme cases, in which the alternative media is either
perfectly reliable $\left( \phi =0\right) $ or completely unreliable $\left(
\phi =1\right) $. If $\phi =0$, there is an equilibrium with accountability,
in which re-election requires not only $m=x$ but also $r=NS$, i.e., the
alternative media doesn't allege that the incumbent is subversive. If $\phi
=1$, there is an equilibrium in which the alternative media is ignored and
re-election is based solely on the mainstream media message, as in the
baseline model.

If the alternative media's reliability is internal, $\phi \in \left(
0,1\right) $, the alternative media's effect is twofold. The first effect
concerns the high-type incumbent's effort. If re-election requires $m=x$ and 
$r=NS$, then increases in $\phi $ demotivate the incumbent, because effort
is only rewarded if the alternative media is truthful. To induce effort
requires $k\leq \left( q-\frac{1}{2}\right) \left( 1-\phi \right) $ or,
equivalently,
\begin{equation*}
\phi \leq \phi _{e}\coloneqq1-\frac{k}{q-\frac{1}{2}}.
\end{equation*}%
The second effect concerns whether the voter listens to the alternative
media. Assuming accountability, the voter removes the incumbent when $m=x$
and $r=S$ if
\begin{equation*}
U_{C}\geq U_{v}\coloneqq1\cdot \Pr \left( \theta _{I}=H|m=x,r=S\right)
-s\cdot \Pr \left( \theta _{I}=S|m=x,r=S\right) =\frac{\pi q\phi -sl}{\pi
q\phi +\left( 1-\pi \right) \frac{1}{2}\phi +l}.
\end{equation*}%
Thus the voter only listens to the alternative media if it is sufficiently
reliable:
\begin{equation*}
\phi \leq \phi _{v}\coloneqq\frac{\left( s+U_{C}\right) l}{\pi q\left(
1-U_{C}\right) -\left( 1-\pi \right) \frac{1}{2}U_{C}}.
\end{equation*}

Combining these effects, we see how $\phi $ affects accountability when $%
U_{C}\in \lbrack \underline{U},\overline{U})$. As shown in Figure 1, if $%
\phi _{v}\leq \phi _{e}$, then if there is accountability in the baseline,
there is also accountability with an alternative media. The voter listens to
the alternative media if $\phi \leq \phi _{v}$ and ignores it otherwise.

\bigskip

\begin{figure}[h]
\centering
\includegraphics[scale=.9]{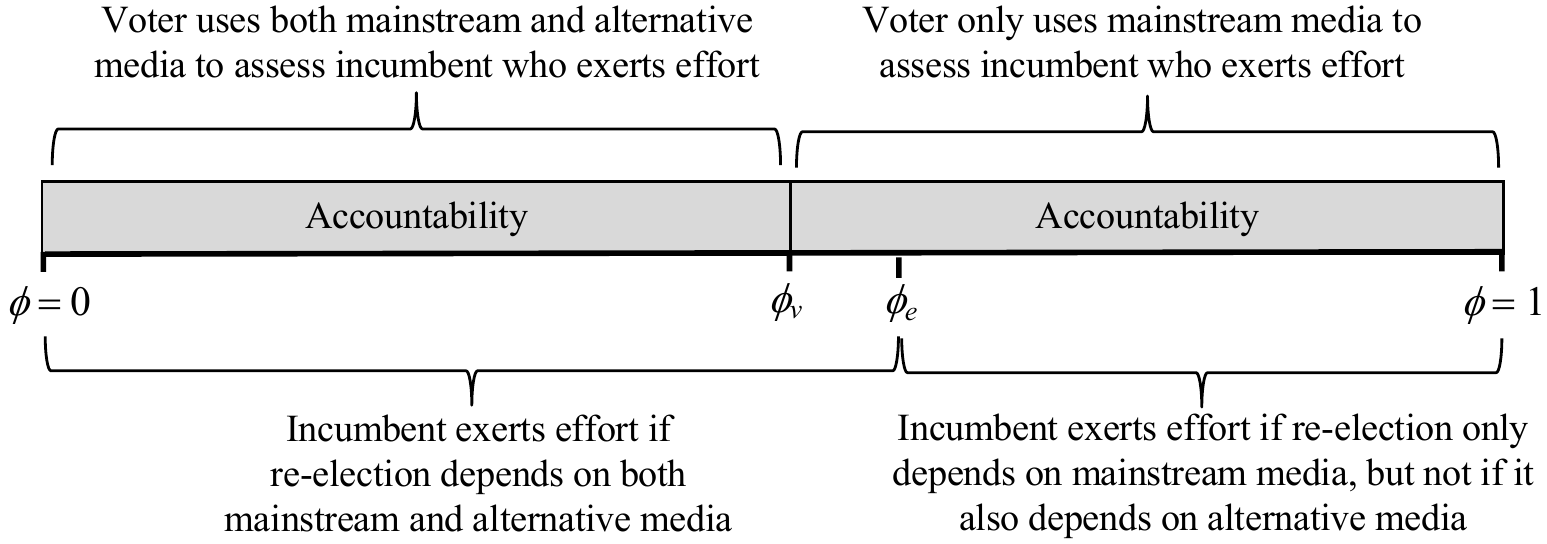}
\caption{Equilibrium as a function of probability that alternative media is
malicious ($\protect\phi $) when effort cost is low ($k<(q-1/2)(1-\protect%
\phi _{v})$) and hence $\protect\phi _{v}<\protect\phi _{e}$.}
\label{figure1}
\end{figure}

If $\phi _{v}>\phi _{e},$ the effect on accountability is more nuanced, as
shown in Figure 2. If $\phi \leq \phi _{e}$ or $\phi >\phi _{v}$, there is
accountability and the voter listens to the alternative media in the first
case and ignores it in the second. But for intermediate reliability $\phi
\in (\phi _{e},\phi _{v}]$, accountability is impossible. This is because
the alternative media is sufficiently reliable that the voter listens to it $%
(\phi \leq \phi _{v})$, but is sufficiently unreliable that the incumbent's
fear of being falsely criticized makes her unwilling to exert effort $\left(
\phi >\phi _{e}\right) $. A necessary condition for the alternative media to
disrupt accountability is $\phi _{v}>\phi _{e}$ or, equivalently $k>\left( q-%
\frac{1}{2}\right) \left( 1-\phi _{v}\right) $, i.e., that policymaking is
difficult, in the sense that it is costly for the incumbent to learn the
correct policy that serves the voter's interest.

\bigskip

\begin{figure}[h]
\centering
\includegraphics[scale=.9]{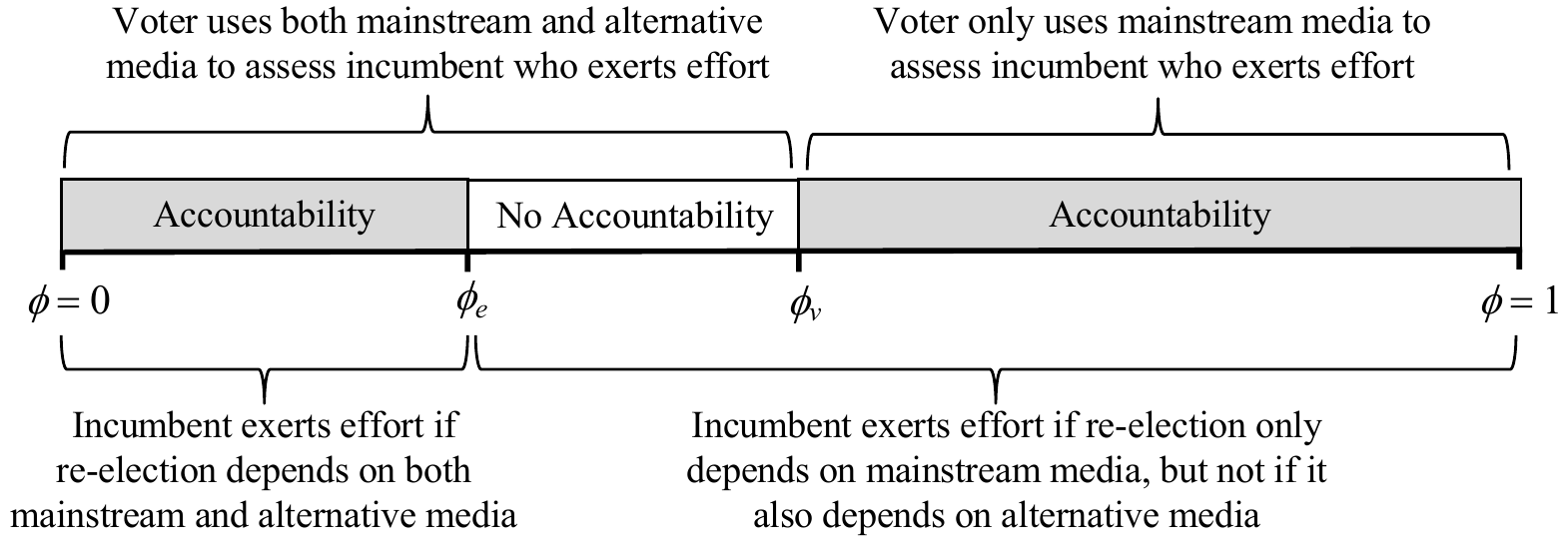}
\caption{Equilibrium as a function of probability that alternative media is
malicious ($\protect\phi $) when effort cost is high ($k>(q-1/2)(1-\protect%
\phi _{v})$) and hence $\protect\phi _{v}>\protect\phi _{e}$.}
\label{figure2}
\end{figure}

In other situations, the alternative media can have positive effects on
accountability. If the challenger is highly appealing $(U_{C}\geq \overline{U%
})$, accountability is impossible in the baseline model and the voter
removes the incumbent even when $m=x.$ The alternative media reveals
additional information, because $r=NS$ means the incumbent is not
subversive. Assuming accountability, the voter's expected utility from
re-electing the incumbent after observing $m=x$ and $r=NS$ is:
\begin{equation*}
\overline{\overline{U}}\coloneqq\Pr \left( \theta _{I}=H|m=x\text{ and }%
r=NS\right) =\frac{\pi q}{\pi q+\left( 1-\pi \right) \frac{1}{2}}>\overline{U%
}.
\end{equation*}
Thus the alternative media makes accountability possible when the challenger
isn't too highly-appealing and false criticism is not so likely as to
demotivate effort, $U_{C}\in \lbrack \overline{U},\overline{\overline{U}})$
and $\phi \leq \phi _{e}$.

\paragraph{Selection}

We next analyze how effectively the voter selects competent types and weeds
out subversive ones, starting with cases where the challenger is moderately
appealing,$\ U_{C}\in \lbrack \underline{U},\overline{U})$.

The case $\phi _{v}\leq \phi _{e}$ (Figure 1) is most straightforward. For $%
\phi \leq \phi _{v}$, there is accountability: the voter listens to both
media outlets and weeds out the subversive type. Local increases in $\phi $
worsen selection of the competent type, who is more frequently falsely
accused and removed from office. For $\phi >\phi _{v}$, there is
accountability, but the alternative media is ignored, and the subversive
type, who induces the mainstream media to report $m=x$, is never weeded out.
In this parameter region, local changes in $\phi $ don't affect the
selection of either type.

The case $\phi _{v}>\phi _{e}$ (Figure 2) is more dramatic, because $\phi $
affects incumbent effort. At $\phi =\phi _{e}$, the equilibrium transitions
from one with accountability to one without accountability, so the voter
loses the benefit of selecting based on the mainstream media message as $%
\phi $ crosses this threshold. Then, as $\phi $ increases above $\phi _{v}$,
accountability is restored, but the alternative media is ignored, so
selection is based solely on the mainstream media message and the subversive
type is never weeded out.

We also note how $\phi $ affects selection based on the mainstream media
message when $U_{C}\notin \lbrack \underline{U},\overline{U})$. With an
unappealing or extremely appealing challenger, $U_{C}<\underline{U}$ or $%
U_{C}>\overline{\overline{U}}$, there is no accountability and the voter
never selects based on the mainstream media message. With a reasonably
highly-appealing challenger, $U_{C}\in \lbrack \overline{U},\overline{%
\overline{U}})$, there is accountability if the alternative media is not
demotivating, $\phi \leq \phi _{e}$, in which case the voter uses
information from both media outlets.

Finally, we note that absent accountability, the voter benefits from
selecting based on the alternative media message if two conditions hold.
First, the incumbent must be sufficiently likely to be a high type to win
re-election when the voter learns that she is non-subversive ($r=NS$) but
learns nothing about her competence. This requires $\pi >U_{C}$. Moreover,
the alternative media must be sufficiently reliable for the voter to remove
the incumbent when $r=S$:
\begin{equation*}
U_{C}\geq U_{a}\coloneqq1\cdot \Pr \left( \theta _{I}=H|r=S\right) -s\cdot
\Pr \left( \theta _{I}=S|r=S\right) =\frac{\pi \phi -sl}{\phi +l},\text{
equivalently }\phi \leq \phi _{a}\coloneqq\frac{\left( s+U_{C}\right) l}{\pi
-U_{C}}.
\end{equation*}

Proposition 2 in the online appendix summarizes equilibria for the model
with alternative media.

\section*{Implications}

We conclude by discussing several implications of our model.%

\paragraph{Failure to heed warnings about democratic backsliding}

Autocratization is a major global trend, in countries such as Brazil,
Hungary, India, Poland, and Turkey. It takes many forms, as incumbents
subvert elections, the bureaucracy, judiciary, and other institutions.
Scholars have identified behind-the-scenes control of the media as one of
the most common forms of backsliding (Bermeo 2016), and evidence on bribes
by President Fujimori of Peru shows that he placed an especially high value
on the media (McMillan and Zoido 2004). Alternative media sometimes serve as
independent information sources and undermine support for autocrats (Knight
and Tribin 2019). But, as noted by Bermeo, they often face a credibility
problem, when rulers accuse them of being special interests, representatives
of a discredited old order, or tools of foreign powers.

In our model, unless the alternative media is seen as being highly reliable $%
\left( \phi <\min \left\{ \phi _{e},\phi _{v}\right\} \right) $, elections
can achieve at most one of two important goals:\ ensuring accountability for
regular incumbents and removing aspiring autocrats (see Figures 1 and 2).
When the alternative media is seen as being sufficiently unreliable $\left(
\phi >\phi _{v}\right) $, voters make electoral decisions based solely on
the mainstream media announcement. This means a subversive incumbent can win
re-election by inducing the mainstream media to praise her policy choices.
Although voters are aware that such praise might just be propaganda, they
think the alternative media is probably \textquotedblleft crying
wolf,\textquotedblright\ so they fail to act on its warnings about the
incumbent's subversion, and thus forgo the opportunity to prevent democratic
backsliding.

\paragraph{Diminished trust in mainstream media}

In our model, the voter's belief about the mainstream media's truthfulness
decreases, from $1-\sigma $ to $\frac{\left( 1-\sigma \right) \left( 1-\phi
\right) }{\left( 1-\sigma \right) \left( 1-\phi \right) +\sigma }$, when the
alternative media alleges that it is acting as an incumbent-controlled
propagandist rather than an independent information provider. This is
broadly consistent with the recent decline in trust in traditional media
outlets at a time when alternative media disparage the mainstream media.%

\paragraph{Fake news and accountability}

Another application of our model is to fake news\ that alleges that an
incumbent is an aspiring autocrat who is in cahoots with the mainstream
media. Theories of fake news include Allcott and Gentzkow (2017), Yea
(2018), and Taylor (2019). While these models analyze the generation of
false claims as well as effects on voter behavior, our model speaks to the
accountability effects of fake news. Specifically, the alternative media is
most beneficial when it is genuine and known to be genuine. In the opposite
extreme, an alternative media that is known to be fake is ignored and does
not disrupt accountability. What is most problematic for accountability is
an alternative media that is too genuine to be ignored yet too fake to
motivate good policymaking.

Our theory may shed light on possible effects of tech giants' recent
measures to battle fake news. In particular, it recommends strong measures
adopted by Youtube and Twitter, such as labeling or removing bogus material,
rather than Facebook's more lenient approach of only removing material that
has been edited via artificial intelligence (Alba and Cogner 2020). In fact,
lenient measures can be worse than doing nothing, if they wind up disrupting
accountability by inadvertently increasing the credibility of fake news that
is not removed from a site. In contrast, strong measures restore voters'
confidence in the materials being posted on the platform and reactivate
their roles in upholding electoral accountability and selection. A
consequence of this is expedited reputation building for truthful
alternative media sources, whose intentions could long remain uncertain and
indistinguishable from those of a myriad of other sources without the
intervention of platform owners.


\section*{Acknowledgements}
We thank Carlo Horz, Floyd Zhang, the editor, three anonymous referees, and audiences at LSE, Warwick, NYU, and the
2019 Comparative Politics and Formal Theory\ conference at UC Berkeley for
helpful comments on this paper.

\newpage
\section*{References}


\noindent Alba, Davey, and Kate Conger. 2020.
\textquotedblleft Twitter Says It Will Police Fake Photos And
Videos.\textquotedblright\ \emph{The New York Times}, February 5, 2020, p.
B4.

\noindent Allcott, Hunt, and Matthew Gentzkow. 2017. \textquotedblleft
Social Media and Fake News in the 2016 Election.\textquotedblright\ \emph{%
Journal of Economic Perspectives }31:211-236.

\noindent Ashworth, Scott, and Kenneth W. Shotts. 2010. \textquotedblleft
Does Informative Media Commentary Reduce Politicians' Incentives to
Pander?\textquotedblright\ \emph{Journal of Public Economics} 94:838-847.

\noindent Bermeo, Nancy. 2016. \textquotedblleft On Democratic
Backsliding.\textquotedblright\ \emph{Journal of Democracy }27:5-19.

\noindent Besley, Timothy, and Andrea Prat. 2006. \textquotedblleft
Handcuffs for the Grabbing Hand? Media Capture and Government
Accountability.\textquotedblright\ \emph{American Economic Review }%
96:720-736.

\noindent Cheah, Hon Foong. 2016. \textquotedblleft Does Foreign Media Entry
Discipline or Provoke Local Media Bias?\textquotedblright\ \emph{%
International Journal of Economic Theory }12:335-359.

\noindent Egorov, Georgy, Sergei Guriev, and Konstantin Sonin. 2009.
\textquotedblleft Why Resource-poor Dictators Allow Freer Media: A Theory
and Evidence from Panel Data.\textquotedblright\ \emph{American Political
Science Review} 103:645-668.

\noindent Gehlbach, Scott, and Konstantin Sonin. 2014. \textquotedblleft
Government Control of the Media.\textquotedblright\ \emph{Journal of Public
Economics} 118:163-171.

\noindent Guriev, Sergei, and Daniel Treisman. 2019. \textquotedblleft
Informational Autocrats.\textquotedblright\ \emph{Journal of Economic
Perspectives} 33(4): 100-127.

\noindent Horz, Carlo M. 2018. \textquotedblleft Propaganda and
Skepticism.\textquotedblright\ Unpublished manuscript.

\noindent Knight, Brian, and Ana Tribin. 2019. \textquotedblleft The Value
of Opposition Media: Evidence from Chavez's Venezuela.\textquotedblright\
Unpublished manuscript.

\noindent Little, Andrew T. 2017. \textquotedblleft Propaganda and
Credulity.\textquotedblright\ \emph{Games and Economic Behavior }102:224-232.

\noindent Lorentzen, Peter. 2014. \textquotedblleft China's Strategic
Censorship.\textquotedblright\ \emph{American Journal of Political Science }%
58:402-414.

\noindent Luo, Zhaotian, and Adam Przeworski. 2019. \textquotedblleft
Democracy and its Vulnerabilities:\ Dynamics of Democratic
Backsliding.\textquotedblright\ Unpublished manuscript.

\noindent McMillan, John, and Pablo Zoido. 2004. \textquotedblleft How to
Subvert Democracy: Montesinos in Peru.\textquotedblright\ \emph{Journal of
Economic Perspectives }18:69-92.

\noindent Nalepa, Monika, Georg Vanberg, and Caterina Chiopris. 2018.
\textquotedblleft Authoritarian Backsliding.\textquotedblright\ Unpublished
manuscript.

\noindent Svolik, Milan W. 2013. \textquotedblleft Learning to Love
Democracy: Electoral Accountability and the Success of
Democracy.\textquotedblright\ \emph{American Journal of Political Science }%
57:685-702.

\noindent Svolik, Milan W. 2020. \textquotedblleft When Polarization Trumps
Civic Virtue: Partisan Conflict and the Subversion of Democracy by
Incumbent.\textquotedblright\ \emph{Quarterly Journal of Political Science }%
15:3-31.

\noindent Taylor, Zachary. 2019. \textquotedblleft Persuasion with Fake
News.\textquotedblright\ Unpublished manuscript.

\noindent Warren, Patrick L. 2012. \textquotedblleft Independent Auditors,
Bias, and Political Agency.\textquotedblright\ \emph{Journal of Public
Economics }96:78-88.

\noindent Wolton, Stephane. 2019. \textquotedblleft Are Biased Media Bad for
Democracy.\textquotedblright\ \emph{American Journal of Political Science }%
63:548-562\emph{.}

\noindent Yea, Sangjun. 2018. \textquotedblleft Persuasion Under the
Influence of Fake News.\textquotedblright\ Unpublished manuscript.

\newpage

\noindent Anqi Li is an Assistant Professor of Economics at the Department of Economics, Washington University in St. Louis, St. Louis, MO 63105.

\bigskip

\noindent Davin Raiha is a Visiting Assistant Professor of Business, Economics, and Public policy at the Kelly School of Business, Indiana University, Bloomington, IN 47405.  

\bigskip

\noindent Kenneth W. Shotts is the David S. and Ann M. Barlow Professor of Political Economy at the Stanford Graduate School of Business, Stanford University, Stanford, CA 94305. 
\newpage

\end{document}